\documentclass[showpacs,amsmath,amssymb,prb,aps,twocolumn]{revtex4}
\usepackage{graphicx}
\begin{document}
\title {On the non-V-representability of currents in time-dependent many
particle systems}
\author{Roberto D'Agosta}
\affiliation{Department of Physics and Astronomy,Ê University of Missouri,
Columbia, Missouri 65211, USA}
\author {Giovanni Vignale}
\email{vignaleg@missouri.edu}
\affiliation{Department of Physics and Astronomy,Ê University of Missouri,
Columbia, Missouri 65211, USA}
\date{\today}

\begin{abstract} We argue that an arbitrarily chosen time-dependent current
density is generically non-V-representable in a many-particle system, i.e., it
cannot be obtained by applying only a time-dependent scalar potential to the
system.  Furthermore we show by a concrete example that even a current that is
V-representable in an interacting many-particle system may (and in general will)
turn out to be non-V-representable when the interaction between the particles is
turned off.
\end{abstract} 
\pacs{} \maketitle

\section {Introduction} 
Since the beginning of density functional theory (DFT)\cite{Hohenberg1964,
Kohn1965,Dreizler1990}, the problem of answering the following questions has
been recognized to be of fundamental importance:
\begin{enumerate}
\item[(1)] Given a (positive) particle density $n(\vec r)$ in an
  N-particle system is there a local potential $V(\vec r)$ that
  produces this density in the ground-state of the system?
\item[(2)] If a certain particle density $n(\vec r)$ arises in the
  ground-state of an N-particle system subjected to a local potential
  $V(\vec r)$,  is there a local potential $V_s(\vec r)$ that
  produces the same density in the ground-state of the same system
  with the particle-particle interactions turned off?
\end{enumerate} 
The first question is known as the {\it V-representability question} and a given
density is said to be {\it V-representable} if the answer is affirmative. The
original formulation of DFT by Hohenberg and Kohn\cite{Hohenberg1964} made heavy
use of the assumption that ``all reasonable densities are V-representable".
Subsequent work\cite{Levy1979} has shown that this assumption is necessary only
when one tries to prove the existence of the functional derivative of the energy
functional. The second question lies at the very heart of the Kohn-Sham
formalism.\cite{Kohn1965} Recall that within this formalism one tries to obtain
the ground-state density of an interacting many-particle system by applying a
local potential to a non-interacting version of the same system. Obviously, it
is vital to the success of this theory that the target density, which is  by
assumption  V-representable, be also non-interacting V-representable -- i.e.,
the answer to question (2) must be affirmative.

Even though mathematically rigorous answers to the two V-representability
questions are not known, DFT has been widely applied to the calculation of the
electronic structure of matter. In these calculations it is tacitly assumed that
the set of V-representable densities in both interacting and non-interacting
systems is dense enough to approximate to an arbitrary level of accuracy any
physical ground-state density (these beliefs are supported by mathematical work
on lattice system\cite{Chayes1985}).  These assumptions have been automatically
transplanted to the relatively younger field of time-dependent density
functional theory (TDDFT)\cite{Runge1984,Trickey1990,Gross1996}, where the
questions are whether a given {\it time-dependent} particle density $n(\vec
r,t)$ evolving from a given initial state $\Psi$ can be produced by a local
time-dependent scalar potential $V(\vec r,t)$ and, in the affirmative case,
whether the same density can also be produced by a time-dependent scalar
potential $V_s(\vec r,t)$ starting from an initial state $\Psi_s$, now with
interactions turned off.

In this paper we are not going to challenge the wisdom of the standard
V-representability assumptions in DFT or TDDFT, but rather examine whether such
assumptions can be plausibly extended to the 
{\it particle current density} $\vec j(\vec r,t)$. There are good reasons to
undertake this study. During the past ten years we have seen many indications
that the time-dependent current
density,\cite{Ghosh1988,Vignale1997,Ullrich2002,Maitra2003,Vignale2004} together
with the initial state of the system (and hence the initial density) may provide
a more fundamental description  of the dynamics. Indeed, the proof that the
time-dependent current density determines the external scalar potential is the
very first step in the proof of the Runge-Gross (RG) theorem -- the foundation
theorem of TDDFT.\cite{Runge1984} 
However, the RG theorem does not say anything about the V-representability
question for the current, i.e. whether a given time-dependent current can be
produced by a local time-dependent scalar potential.  Reasoning by analogy with
the particle density this has led (or, as we are going to show, misled) some
workers to believe that the V-representability assumption for the current
density is about as plausible as the corresponding assumption for the density,
and that therefore any physical current density can be approximated to an
arbitrary degree of precision by the current density generated by suitably
chosen {\it scalar} potential, in an interacting as well as in a non-interacting
system.\cite{Gross1996}

The purpose of this paper is to show that this is not the case. Due to the
vector character of the current density it is usually impossible for an
arbitrary current to be generated by a single scalar function -- the potential. 
Even in those special  (but physically very relevant) cases in which this can be
done, the current will not be simultaneously representable in the
non-interacting system. A more general theory that makes use of an effective
{\it vector} potential to generate the current is therefore needed: such a
theory exists and it is known as time-dependent current density functional
theory (TDCDFT)\cite{Ghosh1988,Vignale2004}.

\section{Non-V-representability of generic current densities} 

In this section we present our main argument against V-representability of the
current density. Recall that the vector field $\vec j(\vec r,t)$, like any other
vector field, can be written as the sum of a longitudinal component $\vec
j_L(\vec r,t)$, a transversal component $\vec j_T(\vec r,t)$, and a constant
that can be assumed to vanish if the full current has to vanish at infinity:
\begin{equation}
\label{jdecomposition}
\vec j(\vec r,t)=\vec j_L(\vec r,t)+\vec j_T(\vec r,t).
\end{equation}
The longitudinal current density $\vec j_L(\vec r,t)$ is curl-free and can
therefore be represented as the gradient of a scalar field, while the
transversal current $\vec j_T(\vec r,t)$ is divergence-free and can therefore be
represented as the curl of a vector field. The spatial Fourier components of
$\vec j_L$ and $\vec j_T$, denoted by $\vec j_L(\vec q,t)$ and $\vec j_T(\vec
q,t)$ respectively, are obtained by projecting the Fourier component of  $\vec
j(\vec q,t)$  along directions parallel and perpendicular to the unit vector
$\hat q$,
\begin{equation}
\vec j_L(\vec q,t) = \left(\vec j(\vec q,t) \cdot \hat q\right)~\hat q,
\end{equation}
and
\begin{equation}
\vec j_T(\vec q,t) = \vec j(\vec q,t)- \vec j_L(\vec q,t).
\end{equation}
Notice that the particle density is completely  determined by the longitudinal
current density since, according to the continuity equation, one has
\begin{equation}
\label{continuity}
\frac{ 
\partial n(\vec r,t)}{ 
\partial t} = - \vec \nabla \cdot \vec j_L(\vec r,t)
\end{equation}
with initial condition $n(\vec r,t)=n_0(\vec r)$.

We begin our argument by assuming that a certain current density $\vec j(\vec
r,t)$ {\it is} V-representable, and let $\vec j_{L}(\vec r,t)$ and $\vec
j_{T}(\vec r,t)$ denote its longitudinal and transverse components,
respectively. According to the RG theorem, the potential $V(\vec r,t)$ that
produces $\vec j (\vec r,t)$ is unique up to an arbitrary function of time.
Consider now a second current density $\vec j' (\vec r,t)=\vec j'_{L}(\vec
r,t)+\vec j'_{T} (\vec r,t)=\vec j_{L}(\vec r,t)+\vec j'_{T} (\vec r,t)$, i.e.
$\vec j'(\vec r,t)$ differs from $\vec j(\vec r,t)$ only because its transverse
component, $\vec j'_{T}(\vec r,t)$ differs from $\vec j_{T}(\vec r,t)$.  We
claim that $j' (\vec r,t)$ is not V-representable.  Indeed, if it were
V-representable, then there would be a potential $V' (\vec r,t) \neq V(\vec
r,t)$ that generates it (here and in the following the $\neq$ sign means that
two potentials differ by more than a mere function of time).  But this is
impossible, since, according to Eq.~(\ref{continuity}), these two different
potentials would give the same particle density, in contradiction with the
Runge-Gross theorem.\cite{Runge1984} Thus, for a given longitudinal current
density $\vec j_L(\vec r,t)$ there is at most one transverse current density
$\vec j_T(\vec r,t)$ such that the full current density $\vec j_L(\vec r,t)+\vec
j_T(\vec r,t)$ is V-representable.

The ease with which, given a V-representable current density, we were able to
construct infinitely many non-V-representable ones is a strong indication that
V-representable currents are a rather exceptional occurrence in the space of all
possible currents.  To strengthen the argument let us make the plausible
assumption that the mapping from potentials to V-representable currents, via the
solution of the time-dependent Schr\"odinger equation, is not only invertible
(RG theorem), but also continuous.  This implies that, within the subset of 
V-representable current densities,  $\vec j_T(\vec r,t)$  is a continuous
functional of $\vec j_L(\vec r,t)$.   Let $\vec j(\vec r,t) = \vec j_L(\vec
r,t)+\vec j_T(\vec r,t)$ be a V-representable current density.  Consider then a
small ``neighborhood" of the non-V-representable current density $\vec
j^\prime(\vec r,t)=\vec j_L(\vec r,t)+\vec j_T'(\vec r,t)$ and let $\vec
j_1(\vec r,t)$ be a (hypothetical) V-representable current density in this
neighborhood.  Since, by choice, the longitudinal component of $\vec j_1$, $\vec
j_{1L}$, is close to $\vec j_{L}$, the continuity of the mapping from $\vec j_L$
to $\vec j_T$ for V-representable currents implies that the transverse component
of $\vec j_1$, $\vec j_{1T}$, is close to $\vec j_{T}$. But, this cannot be true
for a sufficiently small neighborhood of $\vec j^\prime(\vec r,t)$ if the
difference between $\vec j_T$ and $\vec j_T'$ is finite  (see Fig. 1).
\begin{figure}
\includegraphics[width=6cm]{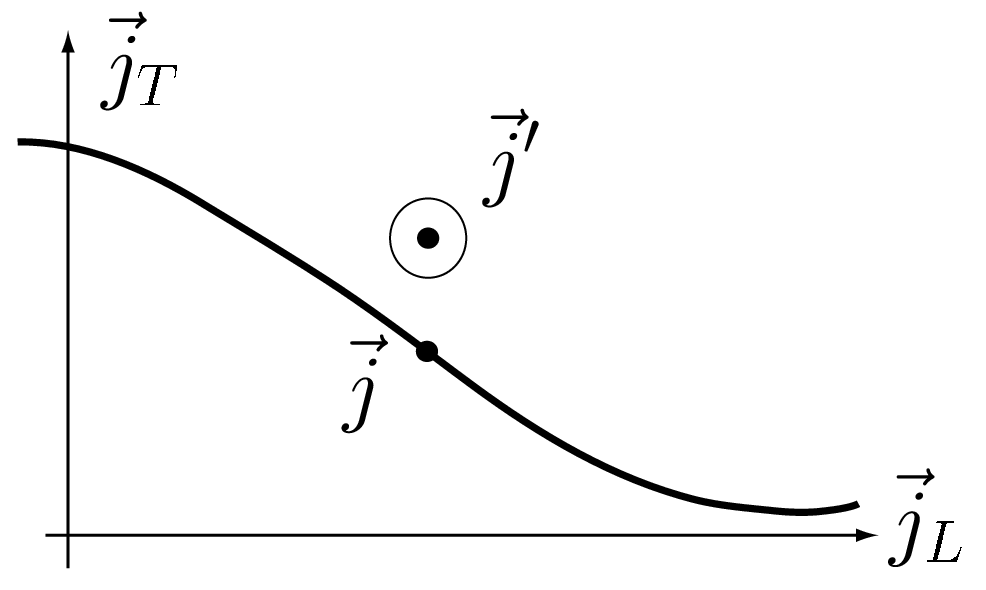}
\caption{A simple cartoon of the sparsity of the V-representable
  current densities. The V-representable current density $\vec j$ lies on a
continuous hypersurface (here schematized as a continuous curve)  in current
density space.  Due to the continuity of the mapping between $\vec j_L$ and
$\vec j_T$ a  sufficiently small neighborhood of the non-V-representable current
density $\vec j'$ contains only non-V-representable current densities.} 
\end{figure} 
We conclude that every non-V-representable current density is surrounded by a
neighborhood that contains only non-V-representable current densities: the set
of V-representable current densities is not ``dense" in the space of all
possible current densities. In Appendix \ref{formal} we give another more
mathematical argument to support our statement.
\section{Interacting  versus non-interacting V-representability} 
Undeterred by the above arguments one might insist that, after all, the task of
the Kohn-Sham theory is to approximate V-representable current densities by
non-interacting V-representable ones. We know that the set of V-representable
current densities is characterized by a certain $\vec j_{T}$ vs $\vec j_{L}$
relation: it is the presence of this constraint that makes the set so ``sparse".
Similarly, in the non-interacting system, the set of the V-representable current
densities is characterized by another $\vec j_{T}$ vs $\vec j_{L}$ relation.
Wouldn't it be nice if these two relations happened to be the same relation, so
that a Kohn-Sham potential yielding the correct density would also automatically
yield the correct current density?

This conjecture has found its way in the literature\cite{Gross1996}, so it is
important to examine it carefully.  In this section we construct an example of a
V-representable current density, which is definitely non-V-representable in the
non-interacting (Kohn-Sham) system.  Thus, in this example the $\vec j_T$ vs
$\vec j_L$ relation of the interacting system turns out to be incompatible with
$\vec j_T$ vs $\vec j_L$ relation of the corresponding non-interacting one.  We
will argue (without proof) that this state of affairs is quite generic for
currents generated by scalar potentials in anisotropic interacting systems.

The system we consider is a two-dimensional interacting electron liquid which is
initially in the ground-state with inhomogeneous density
\begin{equation}
\label{n0} n_0(\vec r)= \bar n(1 + 2 \gamma \cos \vec G \cdot \vec r),
\end{equation}
where $\vec G$ is a two-dimensional vector parallel to the x-axis, $\bar n$ is
the average density, and $\bar n \gamma$ the amplitude of the density
modulation, with $\gamma\ll1$.  This density is produced by the application of
the static potential
\begin{equation}
\label{V0} V_0(\vec r)= \frac{2 \bar n \gamma}{\chi^h(\vec G)} 
\cos \vec G \cdot \vec r
\end{equation} 
to an initially homogeneous electron liquid of density $\bar n$.\footnote{In
this form of the potential some higher order harmonics, e.g. the ones with wave
vectors $\pm 2 \vec G$,  have been neglected.  However, these higher harmonics,
being themselves of order $\gamma^2$, produce  only higher order  corrections
(at least $\gamma^3$) in the subsequent calculations.  For this reason they have
been neglected.}
Here $\chi^h(\vec G)$ is the static density susceptibility of the homogeneous
electron gas of density $\bar n$ at wave vector $\vec G$.\footnote{Here and in
the following the quantities with the superscript $h$ refer to the {\it
homogeneous} system of density $\bar n$.  Their Fourier transforms have a single
wave vector argument.}  
We now apply to this system a time-dependent, periodic scalar potential of the
form
\begin{equation}
\label{perturbingV} V(\vec r,t)  = V e^{i (\vec q \cdot \vec r - \omega t)}
+~c.c.~,
\end{equation}
where $\vec q$ is a two-dimensional wave vector, which we assume to be much
smaller in magnitude than both $G$ and $k_F$ ($\vec{k}_F$ being the Fermi wave
vector corresponding to the average density $\bar n$).  The frequency $\omega$
is   assumed to be larger than both $v_F q$ and $v_F G$, where $v_F$ is the
Fermi velocity associated with $k_F$.

It is well known that the time-dependent potential of Eq.~(\ref{perturbingV})
can be recast as a longitudinal vector potential
\begin{equation}
\label{perturbingA}
\vec A(\vec r,t) = \frac{\vec q V}{\omega} e^{i (\vec q \cdot \vec r -
  \omega t)} +~c.c.~,
\end{equation}
so we can say that the induced current density is
\begin{equation}
\label{inducedcurrent} j_{\alpha}(\vec r,t) = \sum_{l,\beta} \chi_{\alpha
\beta}(\vec q + l \vec G,\vec q,\omega) \frac{q_j V}{\omega} e^{i[(\vec q + l
\vec G) \cdot \vec r - \omega t]}+~c.c.~,
\end{equation}
where $\chi_{\alpha \beta}(\vec q + l \vec G, \vec q,\omega)$ is the dynamical
linear response function of the inhomogeneous liquid, which connects the
cartesian $\beta$ component of the vector potential amplitude at wave vector
$\vec q$ to the cartesian $\alpha$ component of the current density amplitude at
wave vector $\vec q+ l \vec G$, where $l$ is an integer.  It should be noted
that this response function coincides with the homogeneous response function
$\chi^h_{\alpha \beta} (\vec q, \omega)$ up to corrections of order $\gamma ^2$
when $l=0$, and is of first order in $\gamma$ when $l =
\pm 1$.  The components with $|l| \geq 2$ are of order $\gamma^2$ at least.

We now want to show that the exact current of Eq.~(\ref{inducedcurrent}) cannot
be obtained in a non-interacting system subjected only to scalar potentials that
yield the exact density.  We first notice that the ground-state density
$n_0(\vec r)$ is enforced in a non-interacting electron gas by the scalar
potential
\begin{equation}
\label{V0s} V_{0,s}(\vec r)= \frac{2 \bar n \gamma}{\chi^h_0(\vec G)} \cos \vec
G \cdot \vec r~,
\end{equation} 
where $\chi^h_0(\vec G)$ is the static density susceptibility of the
noninteracting electron gas of density $\bar n$ at wave vector $\vec G$. We know
from the invertibility of the mapping between vector potentials and currents
that the exact current density of Eq.~(\ref{inducedcurrent}) can be generated in
the non-interacting electron gas by one and only one time-dependent vector
potential,
\begin{widetext}
\begin{eqnarray}
\label{perturbingAs} A_{s,\alpha}(\vec r,t)  &=& A_\alpha(\vec r,t)+
\sum_{l,l',\beta,\delta} f_{Hxc,\alpha \beta}(\vec q + l \vec G,\vec q + l' \vec
G,\omega) \chi_{\beta \delta}(\vec q + l' \vec G,\omega)
\frac{q_\delta V}{\omega}e^{i[(\vec q + l \vec G) \cdot \vec r -
    \omega t]} +~c.c.~,
 \end{eqnarray}
 where $f_{Hxc,\alpha \beta}(\vec q+l \vec G,\vec q+l' \vec G,\omega)$
 is the sum of the so-called exchange-correlation kernel $f_{xc,\alpha
   \beta}(\vec q+l \vec G,\vec q+l' \vec G,\omega)$ and the Hartree
 kernel $f_{H,\alpha \beta}(\vec q+l \vec G,\vec q+l' \vec G,\omega)
 \equiv \frac{2 \pi e^2}{ |\vec q + l \vec G|}\frac{(q_\alpha + l
   G_\alpha)(q_\beta + l G_\beta)}{\omega^2} \delta_{l l'}$ for our
 system.  Thus, in order to prove that the current of
 Eq.~(\ref{inducedcurrent}) cannot be generated, in the
 non-interacting system, by a scalar potential, we only need to show that
 the vector potential $\vec A_s (\vec r,t)$ has a finite transversal
 component.  For, if this is the case, then the uniqueness of $A_s$
 guarantees that the current density cannot be produced by a purely
 longitudinal vector potential, and hence not by a simple scalar
 potential.
 
 In order to establish the existence of a transversal component of $\vec
 A_s$ we focus on the Fourier component at wave vector $\vec q$ and we
 discard both $\vec A$ and the contribution of the Hartree kernel
 because they are purely longitudinal fields.  The quantity of
 interest is thus the exchange-correlation vector potential
 \begin{equation}
 A_{xc, \alpha}(\vec q,\omega) = \sum_{l,\beta,\delta} f_{xc,\alpha
   \beta}(\vec q,\vec q + l \vec G,\omega) \chi_{\beta\delta}(\vec q +
 l \vec G,\vec q,\omega) \frac{q_\delta V}{\omega}.
 \end{equation}
 The first non-vanishing term in the expansion of this
 quantity in powers of $\gamma$ is of order $\gamma^2$:
 \begin{eqnarray}
\label{Atransverse}
  A_{xc, \alpha}(\vec q,\omega) &\simeq& \sum_{\beta,\delta}
  \left\{f^h_{xc,\alpha \beta}(\vec q,\omega)
  \chi^{(2)}_{\beta\delta}(\vec q,\vec q,\omega)+ f^{(2)}_{xc,\alpha
    \beta}(\vec q,\vec q,\omega) \chi^{h}_{\beta\delta}(\vec q,\omega)
  \right. \nonumber \\
&+& \left.f^{(1)}_{xc,\alpha \beta}(\vec q,\vec
  q+\vec G,\omega) \chi^{(1)}_{\beta\delta}(\vec q + \vec G,\vec
  q,\omega) +f^{(1)}_{xc,\alpha \beta}(\vec q,\vec q-\vec G,\omega)
  \chi^{(1)}_{\beta\delta}(\vec q - \vec G,\vec
  q,\omega)\right\}\frac{q_\delta V}{\omega},
 \end{eqnarray} 
where the quantities with superscripts $^{(1)}$ and $^{(2)}$ refer to the
inhomogeneous sytem of density $n_0(\vec r)$ and are first order and second
order in $\gamma$, respectively.
 
To proceed we now make use of certain exact identities, which can be obtained
starting from identities that were derived 
in Ref. \onlinecite{Vignale1996} from a careful consideration of the behavior of
the current response function and its associated xc kernel under transformation
to an accelerated reference frame. These ``acceleration identities" are
summarized in the following four equations:
 \begin{eqnarray}
\label{accelerationidentity1}
 \chi^{(1)}_{\alpha \beta}(\vec G,\vec 0,\omega) &=& \frac{\bar n
   \gamma}{m} \left [\delta_{\alpha \beta} -
 \frac{\chi^h(G,\omega)}{\chi^h(G)} P^L_{\alpha \beta}(\vec G)
 \right ],\\
\label{accelerationidentity2}
 \chi^{(2)}_{\alpha \beta}(\vec 0,\vec 0,\omega) &=& - 2
 \left(\frac{\gamma\bar n G}{m \omega}\right)^2 \frac{1}{\chi^h(G)}
 \left[1 - \frac{\chi^h(G,\omega)}{\chi^h(G)}\right] P^L_{\alpha
   \beta}(\vec G) ,\\
\label{accelerationidentity3} f^{(1)}_{xc,\alpha \beta}(\vec G,\vec 0,\omega)
&=& -\frac{\gamma
  G^2}{\omega^2} \left\{\left[f^{h}_{xcL}(G,\omega) -
  f^{h}_{xcL}(G,0)\right] P^L_{\alpha \beta}(\vec G)+f^{h}_{xcT}(G,\omega)
P^T_{\alpha \beta}(\vec G)\right\},\\
\label{accelerationidentity4} f^{(2)}_{xc,\alpha \beta}(\vec 0,\vec 0,\omega)
&=& -2 \gamma f^{(1)}_{xc,\alpha \beta}(\vec G,\vec 0,\omega),
\end{eqnarray}
\end{widetext} where $f^{h}_{xcL}(G,\omega)$ and $f^{h}_{xcT}(G,\omega)$ are the
longitudinal and transverse xc kernels of the homogeneous electron liquid at
density $\bar n$ and wave vector $\vec G$, while $P^L_{\alpha \beta}(\vec G)
\equiv \frac{G_\alpha G_\beta}{G^2}$ and $P^T_{\alpha \beta}(\vec G) \equiv
\delta_{\alpha \beta} - P^L_{\alpha
  \beta}(\vec G) $ are the projectors parallel and perpendicular to the
direction of $\vec G$, respectively. The derivation of these identities is
briefly presented in Appendix \ref{acceleration}.

What makes these identities relevant to the evaluation of $\vec A_{xc}(\vec
q,\omega)$ is the fact that in our model $q$ is much smaller than $|\vec G|$ or
$k_F$: therefore the quantities appearing in Eq.~(\ref{Atransverse}) can be
evaluated in the $q \to 0$ limit where they reduce precisely to the quantities
that appear in
Eqs.~(\ref{accelerationidentity1})--(\ref{accelerationidentity4}). Underlying
the calculation is of course the assumption that the $q \to 0$ limits of the
current response functions and xc kernels are regular -- an assumption we have
presently no reason to doubt.

By making use of
Eqs.~(\ref{accelerationidentity1})--(\ref{accelerationidentity4}) in
Eq.~(\ref{Atransverse}), and by discarding all but the leading-order terms in
$q$ we arrive, after some algebra, at
 \begin{eqnarray}
  \vec A_{xc}(\vec q,\omega) & = &2\gamma^2 \frac{\bar n \vec G (\vec
    G \cdot \vec q)}{m \omega^2} \frac{\chi^h(G,\omega)}{\chi^h(G)}
  \nonumber\\
&\times&\left[f^{h}_{xcL}(G,\omega) -
    f^{h}_{xcL}(G,0)\right] \frac{V}{\omega}~.
 \end{eqnarray}
 This vector has a component perpendicular to $\vec q$, unless $\vec
 q$ happens to be either parallel or perpendicular to $\vec G$:
 choosing $\vec q$ in any other direction provides the desired example
 of non V-representability.
 
It is also interesting to ask what is the behavior of the Fourier components of
$\vec A_{xc}$ at wave vectors $\vec q \pm \vec G$. These components are first
order in $\gamma$ and one might think that they lead more directly to the
desired result.
 Remarkably, this is not the case: A calculation very similar to the one
described in the previous paragraphs reveals that these components are purely
longitudinal, i.e., parallel to $\vec q \pm \vec G$.  One needs to go to at
least second order in $\gamma$ to see a transverse component of $\vec A_{xc}$.

\section {Discussion} The example worked out in the previous section shows that
a perfectly legitimate V-representable current density can turn out to be
non-V-representable in the non-interacting system.  We believe that this state
of affairs is generic.  Only in exceptional cases will the current density be
V-representable in both the interacting and the non-interacting versions of the
same system.  Hence, in general, the Kohn-Sham equation does not give the
correct value of the transverse current density.

We may now ask, how big an error does one make if one insists on calculating the
transverse current by means of the Kohn-Sham theory? Going back to our model
system it is not difficult to see that the Fourier components of the Kohn-Sham
potential are given by 
\begin{equation}
\label{VKSmodel} V_{KS}(\vec q + l \vec G,\omega) = \frac{\omega}{|\vec q + l
\vec G|} |\vec A_{s,L}(\vec q + l\vec G,\omega)|~,
\end{equation}
i.e., the Kohn-Sham potential is simply the scalar representation of the
longitudinal part of the vector potential $\vec A_s$ calculated in the previous
section ($\vec A_{s,L}$ and $\vec A_{s,T}$ are the longitudinal and transversal
component of $\vec A_s$, 
respectively).  The above equation is accurate up to corrections of order
$\gamma^3$. 
Indeed, because $\vec A_{s,T}$ is of order $\gamma^2$ its influence on the
longitudinal current begins at order $\gamma^3$, implying that $\vec A_{s,L}$
alone fully accounts for the longitudinal current density up to corrections of
order $\gamma^3$. On the other hand, because $V_{KS}$ is equivalent to $\vec
A_{s,L}$ it clearly fails to produce the part of the transverse current that is
due to $\vec A_{s,T}$.  This is of order $\gamma^3$ for the Fourier components
at wave vector $\vec q + l\vec G$ with $l \neq 0$ and of order $\gamma^2$ for
the Fourier component at wave vector $\vec q$.  We conclude that the error on
the transverse current is overall of order $\gamma^2$: this may partly explain
the difficulty of finding examples in which the Kohn-Sham current density
differs significantly from the exact one.

Where does this leave us as to the application of the time-dependent Kohn-Sham
theory to the calculation of current densities?  From a fundamental standpoint
it is clear that only the time-dependent CDFT can provide the right answer.  In
time-dependent CDFT one does not need V-representability, but only the much
weaker $\vec A$-representability assumption.  We know that this assumption holds
true in linear response theory, and it is highly reasonable to assume that the
set of $\vec A$-representable current densities is dense in the space of all
current densities.  On the other hand, we have also found that the error
entailed by the use of the ordinary Kohn-Sham equation of TDDFT is of second
order in the parameter that measures the strength of the density non-uniformity,
and may perhaps be reasonably neglected in practical implementations of the
theory that are based on the local density approximation.

\acknowledgments The authors acknowledge support from NSF Grant No. DMR-0313681.
GV thanks Hardy Gross and Ilya Tokatly for several discussions on the subject of
this paper.

\appendix

\section{A formal proof of the non-dense V-reprensentable current density
set}\label{formal} 

In this section we discuss a different proof of the fact that the
V-representable current density set is not dense in the set of physical
currents. Let us recall that, mathematically, a subset $X$ of the set $A$ is
dense in $A$ if and only if every element of $A$ is the limit of a sequence of
elements in $X$.\cite{Wade} It is easy to see that if $X\not = A$ and $X$ is
closed, i.e. any sequence of elements in $X$ has limit in $X$, then $X$ cannot
be dense in $A$.

Our first observation is that the physical currents, which are integrable and
differentiable together with their derivatives, form therefore a Sobolev
space.\cite{Lieb} Let us now construct a sequence $\{j_{l}\}$ of V-representable
current densities and let us study its limit $l\to\infty$. Because the Sobolev
space is closed, the limit $\vec j_{\infty}(\vec r,t)=\lim_{l\to\infty}\vec
j_{l}(\vec r,t)$ is still an integrable and differentiable function, i.e.,  it
is still a physical current. 
This is also true for the corresponding densities $\{n_{l}\}$ and 
wave-functions $\{\Psi_{l}\}$. We then consider the sequence of potentials
$\{V_{l}\}$ that generates the sequence of current densities $\{j_{l}\}$. The
hypothesis of the continuity of the mapping between potentials and currents
formulated in the text ensures the existence of the limit
\begin{equation}
\lim_{l\to\infty}V_{l}(\vec r,t)=V_{\infty}(\vec r,t).
\end{equation}
We use this result in the sequence of Schr\"odinger equations
\begin{equation}
i
\partial_t \Psi_l=H_l \Psi_l,
\end{equation}
where $H_l$ is defined by the potential $V_l$, to pass to the $l\to \infty$
limit and arrive at the conclusion that $j_\infty$ is V-representable, and the
potential which defines it is $V_\infty$. 
Thus all the sequence of V-representable currents have a V-representable limit,
i.e. this set is closed. Because we proved that the set of non-V-representable
currents is not empty, the V-representable current densities do not form a dense
subset of all the physical current densities.

\section{Acceleration identities for the current response functions of a weakly
inhomogeneous electron liquid}\label{acceleration} Our starting point is the
acceleration identity in real space
\begin{equation}
\label{chistart}
\sum_\delta \int \chi_{\alpha\delta}(\vec r,\vec
r',\omega)\left[\delta_{\delta\beta}+\frac{\nabla'_\delta\nabla'_\beta
    V_0(\vec r')}{m (i\omega)^2}\right]d\vec r' = \frac{n_0(\vec
  r)}{m}\delta_{\alpha\beta},
\end{equation}
which was derived in Ref. \onlinecite{Vignale1996}, beginning on page 205. Both
sides of this identity can be expanded in a series of $\gamma$: The Fourier
transform of the coefficients of this expansion will give us the identities 
(\ref{accelerationidentity1})-(\ref{accelerationidentity2}). 
The first order in $\gamma$ evaluated at wave vector $\vec G$ gives:
\begin{equation}
 \chi^{(1)}_{\alpha \beta}(\vec G,\vec 0,\omega) + \sum_\delta
 \chi^h_{\alpha\delta}(\vec G, \omega) \frac{G_\delta
   G_\beta}{m\omega^2}V_0(\vec G) = \frac{\bar n
   \gamma}{m}\delta_{\alpha\beta}.
 \end{equation}
 Inserting
 \begin{equation}
\label{V0G}
 V_0(\vec G) = \frac{\bar n \gamma}{\chi^h(\vec G)}
 \end{equation} 
 from Eq.(\ref{V0}), and recalling that
 \begin{equation}
 \sum_\delta \chi^h_{\alpha\delta}(\vec G, \omega) G_\delta G_\beta =
 \chi^h(\vec G, \omega) \omega^2 P^L_{\alpha \beta}(\vec G)~,
 \end{equation}
where $\chi^h(\vec G,\omega)$ is the density-density response function, we
arrive at Eq.~(\ref{accelerationidentity1}).

To derive Eq.~(\ref{accelerationidentity2}) we take the $q=0$ component of both
sides of Eq.~(\ref{chistart}) of the second order in $\gamma$.  Since $V_0(\vec
r)$ and $n_0(\vec r)$ are given by Eqs.~(\ref{n0}) and (\ref{V0}) up to
corrections of order $\gamma^3$ we readily obtain
\begin{eqnarray}
\chi^{(2)}_{\alpha\beta}(\vec 0, \vec 0,\omega) &=&-\sum_{\delta}
\left[\chi^{(1)}_{\alpha\delta}(\vec 0, \vec
  G,\omega)+\chi^{(1)}_{\alpha\delta}(\vec 0, -\vec
  G,\omega)\right]\nonumber \\
&\times&\frac{G_\delta
  G_\beta}{m\omega^2}V_0(\vec G).
\end{eqnarray}
Inserting Eq.~(14) for $\chi^{(1)}_{\alpha\delta}(\vec 0, \vec G,\omega)$ and
Eq.~(\ref{V0G}) we immediately arrive at Eq. 
(\ref{accelerationidentity2}).

We proceed similarly for the last two identities,
Eqs.~(\ref{accelerationidentity3}) and~(\ref{accelerationidentity4}). The
starting point in this case is Eq. (27) of Ref. \onlinecite{Vignale1996}:
\begin{equation}
\label{fxcstart}
\int f_{xc,\alpha\beta}(\vec r,\vec r',\omega)n_0(\vec r') d \vec r' = -
\frac{\nabla_\alpha \nabla_\beta V_{s,xc}(\vec r)}{\omega^2}~,
\end{equation}
where, to the required accuracy\footnotemark[17]
\begin{equation}
V_{s,xc}(\vec r) = 2 \bar n \gamma f^h_{xcL}(\vec G,0) \cos \vec G
\cdot \vec r
\end{equation}
is the exchange-correlation part of the static Kohn-Sham potential $V_s(\vec
r)$.  Notice that $f^h_{xcL}(\vec G,0)$ is the {\it scalar} exchange-correlation
kernel of the homogeneous electron liquid at zero frequency, quite different
from the tensorial and frequency-dependent exchange-correlation kernel of the
inhomogeneous system, which appears on the left hand side of
Eq.~(\ref{fxcstart}). Taking the Fourier component at wave vector $\vec G$ of
both sides of Eq.~(\ref{fxcstart}) to first order in $\gamma$, and recalling
that
\begin{eqnarray}
f^h_{xc,\alpha \beta}(\vec G,\omega)& = &\frac{G^2}{\omega^2}
\left[f^h_{xcL}(\vec G,\omega) P^L_{\alpha\beta}(\vec
  G)\right.\nonumber\\
&+&\left. f^h_{xcT}(\vec G,\omega)
  P^T_{\alpha\beta}(\vec G) \right]
\end{eqnarray}
we arrive at Eq.~(\ref{accelerationidentity3}).  Taking the Fourier component of
both sides of Eq.~(\ref{fxcstart}) at wave vector $0$ to second order in
$\gamma$ we finally arrive at Eq.~(\ref{accelerationidentity4}).

\bibliography{dft-biblio,general}

\end{document}